\documentclass[a4paper,12pt]{article}
\usepackage{epsfig,psfrag}
\usepackage{citesort}
\date{\today}
 \normalsize

\newcommand{\bmat}{\left(\begin{array}}
\newcommand{\emat}{\end{array}\right)}
\newcommand{\be}{\begin{equation}}
\newcommand{\ee}{\end{equation}}
\newcommand{\bea}{\begin{eqnarray}}
\newcommand{\eea}{\end{eqnarray}}

\def\lsim{\raise0.3ex\hbox{$\;<$\kern-0.75em\raise-1.1ex\hbox{$\sim\;$}}}
\def\gsim{\raise0.3ex\hbox{$\;>$\kern-0.75em\raise-1.1ex\hbox{$\sim\;$}}}
\def\Frac#1#2{\frac{\displaystyle{#1}}{\displaystyle{#2}}}

\def\jp{J/\psi}
\def\jpp{B_s \to \jp \phi}
\def\jpk{B_d \to J/\psi K_S}
\def\dG{\Delta\Gamma_{s}}
\def\dM{\Delta M_s}
\def\sm{\mbox{\tiny SM}}

\begin{document}
\begin{titlepage}
\renewcommand{\thefootnote}{\fnsymbol{footnote}}
\rightline{IPPP-03-61} \rightline{DCPT-03-122} \rightline{UCL-IPT-03-18}
\vspace{.3cm}
{\Large
\begin{center}
{\bf $B_s^0 - \overline{B}_s^0$ Mixing and the $B_s \to J/\psi \phi$
  Asymmetry in Supersymmetric Models}
\end{center}}
\vspace{.3cm}

\begin{center}
P. Ball$^{1}$,  S. Khalil$^{1,2}$ and E. Kou$^{1,3}$\\
\vspace{.3cm}
$^1$\emph{IPPP, Physics Department, University of Durham, DH1 3LE,
Durham, UK}
\\
$^2$ \emph{Ain Shams University, Faculty of Science, Cairo, 11566,
Egypt}
\\
$^3$ \emph{Institut de Physique Th\'{e}orique, Universit\'{e} Catholique de 
Louvain, Chemin Cyclotron~2, B-1348 Louvain-la-Neuve, Belgium}

\vskip2cm


\end{center}

\vspace{.3cm}
\begin{center}
\small{\bf Abstract}\\[3mm]
\end{center}
We analyse
supersymmetric contributions to $B_s$
mixing  and their impact on mixing-induced 
CP asymmetries, using
the mass insertion approximation.
We discuss in particular the correlation of SUSY effects in 
the CP asymmetries of $B_s\to J/\psi \phi$ 
and $B_d\to \phi K_S$ and find that the  mass insertions  
dominant in $B_s$ mixing and $B_d\to \phi K_S$ are 
$(\delta_{23}^d)_{LL, RR}$ and $(\delta_{23}^d)_{LR, RL}$, respectively.  
We show that models with dominant $(\delta_{23}^d)_{LR, RL}$ can 
accomodate a negative value of $S_{\phi K_S}$, in agreement with 
the Belle measurement of that observable, 
but yield a $B_s$ mixing phase too small
to be observed. On the other hand, models with dominant 
$(\delta_{23}^d)_{LL, RR}$ predict sizeable SUSY 
contributions to both $\dM$ and the mixing phase, but do
not allow the asymmetry in $B_d\to \phi K_S$ 
to become negative, except for small values of
the average down squark mass, which, in turn, entail a value of $\dM$ 
too large to be observed at the Tevatron and the LHC. 
We conclude that 
the observation of $B_s$ mixing at hadron machines, 
together with the confirmation of a negative value of 
$S_{\phi K_S}$, disfavours models with a single dominant mass insertion.
\begin{minipage}[h]{14.0cm}
\end{minipage}
\vskip 0.3cm 
\vfill

\end{titlepage}
%
\section{{\large \bf Introduction}}
The impressive performance of the $B$ factory experiments BaBar and Belle
provides the basis for scrutinizing tests of 
the standard model (SM) picture of flavour structure and 
CP violation in the quark sector, and opens the possibility to probe 
virtual effects from new physics at low energies.
In the supersymmetric extension of the SM, 
a new source of flavour violation  arises from the fact that, in
general, the rotation that translates flavour eigenstates into
mass eigenstates will not be the same for quark and squark fields, which
implies the appearance of a new squark mixing matrix or,
alternatively, that of off-diagonal squark mass terms in a basis where
the quarks are mass-eigenstates and 
both quark and squark fields have undergone the same rotation -- the
so-called super-CKM basis.
A convenient tool for studying the impact
of this new source of flavour violation is the
mass-insertion approximation (MIA), which was first introduced in
\cite{MIA} and since then has been widely used as a largely
model-independent tool for analysing and constraining SUSY effects in
$B$ physics. In the super-CKM basis the couplings of fermions and their
SUSY partners to neutral gauginos are flavour-diagonal and 
flavour-violating SUSY effects are encoded in the nondiagonal entries of
the sfermion mass matrix. The sfermion propagators are expanded
in a series in $\delta = \Delta^2/\tilde m^2_{\tilde q}$, where $\Delta^2$ are
the off-diagonal entries and $\tilde m_{\tilde q}$ is the average sfermion
mass. We assume $\Delta^2\ll \tilde m^2_{\tilde q}$, so that the first term
in the expansion is sufficient, and also that the diagonal sfermion
masses are nearly degenerate.

Flavour-changing box and penguin
processes as observed at the $B$ factories 
are very sensitive to flavour-violating effects beyond the SM, 
and the constraints on or measurement of
nondiagonal squark masses will help to discriminate among various soft
SUSY breaking mechanisms.  
In summer 2002, BaBar and Belle reported the first measurements of 
the mixing-induced CP asymmetry $S_{\phi K_S}$ in $B_d\to \phi K_S$, 
which at the quark level is $b\to s\bar s s$ and thus 
a pure penguin process, which is expected to exhibit, in the SM, the same
mixing-induced CP 
asymmetry as observed in  $\jpk$ \cite{Nir}. 
The experimental results, however, updated in summer 2003, paint a
slightly different picture:
\bea
S_{J/\psi K_S} &=& \phantom{-}0.736 \pm 0.049\quad \mbox{~(BaBar \&
  Belle})~\cite{BaBar1,Belle1}\\[5pt]
S_{\phi K_S} &\stackrel{2002}{=}& -0.39\phantom{6} \pm 0.41~
\cite{BaBar2,Belle2}~\stackrel{2003}{=}~\left\{
\begin{array}{l@{\quad}l}
-0.96 \pm 0.50^{+0.09}_{-0.11}&\mbox{Belle~~~}\cite{Belle3} \\
+0.45 \pm 0.43 \pm 0.07 &\mbox{BaBar~}\cite{BaBar3}
\end{array}
\right.
\eea
Although the experimental situation in $B_d\to \phi K_S$ is not yet 
conclusive, the deviation of $S_{\phi K_S}$ from $S_{J/\psi K_S}$ 
may constitute a first potential glimpse at physics
beyond the SM, and it is both worthwile and timely to pursue any
interpretion of these results in terms of new physics and to analyse
their impact on future measurements to be performed at the $B$
factories or at the Tevatron and the LHC, see e.g.\ 
\cite{KK-phk,KK-etak,BdBd-th1,phiks-th}. 

In the framework of MIA, the measurement of $S_{J/\psi K_S}$, which is
in agreement with the SM expectation, indicates
that $(\delta_{13}^d)_{AB}$, $A,B=L,R$, is small \cite{BdBd-th0},
whereas the result for $S_{\phi K_S}$ indicates 
a relatively large $(\delta_{23}^d)_{AB}$. Furthermore, by including the 
constraints on $(\delta_{23}^d)_{AB}$ from $b\to s\gamma$, 
it was found  \cite{KK-phk} that, for average 
squark masses of order $500\,$GeV, only models with dominant 
$(\delta_{23}^d)_{LR, RL}$ can 
accomodate a negative value of $S_{\phi K_S}$.

$\delta^d_{23}$ insertions also determine the size of SUSY
contributions to $B_s$ mixing and, as a consequence, the
mixing-induced CP asymmetries in  tree-level dominated decays
like e.g.\ $B_s\to J/\psi \phi$, which is one of the benchmark
channels to be studied at hadron machines. Within the SM, the
$B_s$ mixing phase is very small, and consequently $S_{J/\psi \phi}$
expected to be of ${\cal O}(10^{-2})$. 
In SUSY, on the other hand, 
the third-to-second generation ($b\to s$) box diagram 
may carry a sizeable CP violating phase, which   
is described in terms of the same mass insertion 
$(\delta_{23}^d)_{AB}$ governing the CP asymmetry $S_{\phi K_S}$. 
It is therefore both important and instructive to analyse 
all $b\to s$ transitions in the same framework, paying
particular attention to the correlations between observables.
This is the subject of this paper.

Our paper is organised as follows: in Section 2, we recall the master
formulas determining $B_s$ mixing and  the CP asymmetry in $\jpp$ and
discuss the SM expectations for the
$B_s$ mixing parameters and the experimental reach for
 $B_s$ mixing at hadron colliders.
In Section 3, we discuss the dominant SUSY contributions 
to $B_s$ mixing in the framework of the mass insertion approximation. 
In Section 4, we present numerical results and discuss the correlation
between the constraints from $b\to s\gamma$ and $S_{\phi K_S}$, obtained 
previously in Ref.~\cite{KK-phk}, and $B_s$ mixing. 
Section 5 contains a summary and 
conclusions. 

%
\section{{\large \bf\boldmath $B_s$ Mixing and the 
Mixing-Induced CP Asymmetry in $\jpp$}}
\subsection{Master Formulas and New Physics Effects}

Let us begin by recalling 
\footnote{Here we use the convention 
$|B_{s}\rangle_1 =p|B_s^0\rangle +q|\overline{B}_s^0\rangle$ and 
$|B_{s}\rangle_2 =p|B_s^0\rangle -q|\overline{B}_s^0\rangle$ 
where we define {\bf CP}$|P \rangle = + |P\rangle$
and $\dM =M_2-M_1$ and 
$\dG=\Gamma_1-\Gamma_2$.}
the master formulas for $B_s$ mixing and the
resulting mixing-induced asymmetry in $\jpp$.
Like for $B_d$, the mixing angles $p$ and $q$ between the flavour and
mass eigenstates in the $B_s$ system can be expressed in terms
of the $B^0_s-\bar B^0_s$ transition matrix element $M_{12}$:
\be
\frac{q}{p}=\sqrt{\frac{M_{12}^*}{M_{12}}},   \label{eq:11}
\ee
where we have used $\dG \ll \dM$ and $\dG \ll \Gamma^{tot}_{s}$.
The resulting mass and width differences between mass eigenstates are
given by
\bea
\dM&=&-2 M_{12}, \quad
\dG=2\Gamma_{12}\cos\zeta_B,\label{eq:13}
\eea
where $\zeta_B\equiv \arg(\Gamma_{12}/M_{12})$. 
$\Gamma_{12}$ can be computed from diagrams with two insertions
of the $\Delta B = 1$ Hamiltonian and is dominated by the tree
contribution. SUSY effects are very small, so to very good accuracy one can set
\be
\Gamma_{12}=\Gamma_{12}^{\sm}. 
\ee
In the SM, $M_{12}$ is dominated by top quark exchange; 
the mixing phase is given by
\be
{\rm arg}\, M_{12}^{\rm SM} = 2 {\rm arg}\,(
V_{tb}V_{ts}^*) = -2\lambda^2\eta = {\cal O}(10^{-2}). 
\label{eq:15}
\ee
In SUSY, there are new contributions to $M_{12}$ induced by e.g.\
gluino and chargino box diagrams, which potentially carry a large
phase and which we parametrise as
\be
\sqrt{\frac{M_{12}}{M_{12}^{\sm}}}\equiv r_s e^{i\beta_s}, \label{eq:17}
\ee
which entails 
\be\label{xyz}
\dM =r_s^2 \dM^{\sm}, \qquad \dG \simeq \dG^{\sm}\cos
2\beta_s,
\ee
assuming $\beta_s\gg \arg M_{12}^{\sm}$.
The above 
result implies that new physics contributions will always lead to a
decrease of $\dG$,
as was first discussed in Ref.~\cite{Gros}.

Let us now discuss the effect of SUSY on the mixing-induced CP
asymmetry in the tree-dominated decay $B_s\to J/\psi \phi$, which is
expected to be very small in the SM and hence highly susceptible to
large or even moderate new CP violating phases. Although
the final state $\jp \phi$ is not a CP eigenstate, 
but a superposition of CP odd and even states which can be
disentangled by an angular analysis of their decay products 
\cite{DDLR,jpph-th}, the advantage of that channel over the
similar process $B_s\to J/\psi\eta(')$ is the comparatively
clean, although still challenging 
reconstruction of the $\phi$ via $\phi\to K^+K^-$, whereas the $\eta(')$ is
even more elusive. Once the CP-waves have been identified,
the analysis of $\jpp$ proceeds largely along the same lines as that of
$\jpk$, except for the fact that, in contrast to $B_d$ mixing, 
the width difference $\dG$
cannot be neglected and entails a slight modification of the formula
for the asymmetry. Without a separation of the final state CP-waves,
the mixing asymmetry still
depends on hadronic parameters describing the  polarisation 
amplitudes $A_{0,\parallel,\perp}$
characteristic for the
final state ($A_{0,\parallel}$ for CP-even and $A_\perp$ for CP-odd). 
One finds, assuming no direct CP-violation, 
\bea
S_{\jp \phi}\,\sin\dM t
&=& \frac{\Gamma (\overline{B}^0_s\to \jp \phi)-\Gamma (B^0_s\to \jp \phi)}
{\Gamma (\overline{B}^0_s\to \jp \phi)+\Gamma (B^0_s\to \jp \phi)}\nonumber \\
&=&\frac{D\ \mbox{Im}\left[\frac{q}{p}\overline{\rho}_{\rm odd}\right]
+\mbox{Im}\left[\frac{q}{p}\overline{\rho}_{\rm even}\right]}
{D\ F_{\rm odd}(t)+F_{\rm even}(t)} 
\sin\dM t
 \label{eq:1}
\eea
where 
\be
F_{\rm odd, even}(t)=\cosh\left(\frac{\dG}{2}t\right)
+\mbox{Re}\left[\frac{q}{p}\overline{\rho}_{\rm odd, even}\right]
\sinh\left(\frac{\dG}{2}t\right)
\ee
and $D$ encodes the polarisation amplitudes:
\be
D\equiv \frac{|A_{\perp}|^2}{|A_{\parallel}|^2+|A_{0}|^2}.
\ee
$D$, as a hadronic quantity, comes with a certain theoretical
uncertainty. Ref.~\cite{BF}, for instance, quotes $D\approx 0.3\pm
0.2$.
 
The parameter $\overline{\rho}$  is defined as 
\be
\overline{\rho}_{\rm odd, even}
=\frac{A(\overline{B}^0_s\to \jp \phi)_{\rm odd, even}}
{A(B^0_s\to \jp \phi)_{\rm odd, even}}
\ee
and can be computed from the $\Delta B=1$ effective Hamiltonian,
yielding
\be
\overline{\rho}_{\rm odd, even}= \mp
\frac{V_{cb}V_{cs}^*}{V_{cb}^*V_{cs}}=\xi_{\rm odd, even} 
\ee 
with $\xi_{\rm even}=+1$ and $\xi_{\rm odd}=-1$. Accordingly, we have 
\be
\frac{q}{p}\overline{\rho}_{\rm odd, even}\simeq\xi_{\rm odd, even}
e^{-2i\beta_s}. 
\ee

\subsection{Estimate of $\dM^{\sm}$ and $\dG^{\sm}$}
\begin{figure}[t]\vspace{-0.5cm}\begin{center}
\psfrag{ex}[l][l][0.8]{Current exp.\ bound $>$ 13 ps$^{-1}$}
\psfrag{dMs}[l][l][1]{$\dM^{\sm}\  {\rm ps}^{-1}$}
\psfrag{gamma}[l][l]{$\gamma^{\sm}$}
\psfrag{g}[l][l][0.75]{$\gamma$}
\psfrag{b}[l][l][0.75]{$\beta$}
\psfrag{Vub}[l][l][0.75]{$V_{ub}$}
\psfrag{max}[l][l][0.8]{$\dM^{\sm}$ max.}
\psfrag{min}[l][l][0.8]{$\dM^{\sm}$ min.}
\psfrag{maximum}[l][l]{$\dM^{\sm}$ maximum}
\psfrag{minimum}[l][l]{$\dM^{\sm}$ minimum}
\psfrag{rho}[l][l]{$\overline{\rho}$}
\psfrag{eta}[l][l]{$\overline{\eta}$}
\includegraphics[width=16cm]{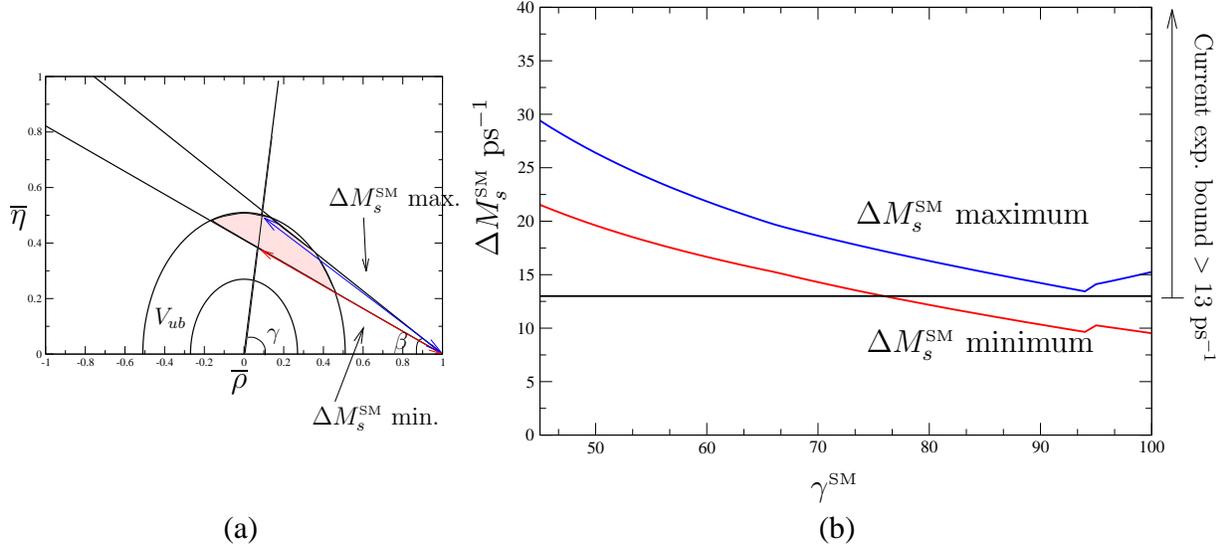}
\caption{(a) Allowed region (shaded area) 
for the apex of the SM unitarity triangle,
  using the constraints from $|V_{ub}/V_{cb}|$ and $\sin 2\beta$. (b)
  $\dM^{\sm}$ as function of $\gamma^{\sm}$ as determined from (a).}
\label{fig:1}
\vspace{-0.5cm}
\end{center}\end{figure}

In order to estimate $\dM^{\sm}$, one usually uses the ratio 
$\dM^{\sm}/\Delta M_d^{\sm}$, in which 
all short-distance effects cancel: 
\be
\frac{\dM^{\sm}}{\Delta M_d^{\sm}}=\frac{M_{B_s}}{M_{B_d}}
\frac{B_{B_s}f^2_{B_{s}}}{B_{B_d}f^2_{B_{d}}}\frac{|V_{ts}|^2}{|V_{td}|^2}. 
\label{eq:24-1}
\ee
The remaining ratio of hadronic parameters has been calculated on
the lattice yielding \cite{lat1}
$$\frac{B_{B_s}(m_b)f^2_{B_{s}}}{B_{B_d}(m_b)f^2_{B_{d}}} =
(1.15\pm 0.06 ^{+0.07}_{-0.00})^2,$$
where the asymmetric error is due to the effect of chiral logarithms in the
quenched approximation. 
In many SUSY models the dominant new contributions to $B_d$ mixing
involve transitions between the third and the first generation and are
thus suppressed by the corresponding CKM matrix elements, so that
$B_d$ mixing is saturated by 
the SM contribution 
\cite{GK,BdBd-th0,BdBd-th1,BdBd-th2} and we can
 assume 
$\Delta M_d=\Delta M_d^{\sm}$. 
$\Delta M_d$ is measured from the time-dependence 
of $B_d$ mixing and is rather precisely known  \cite{PDG}:
$$(\Delta M_d)_{\rm exp}=(0.489\pm0.008)\,{\rm ps}^{-1}.$$ 
As for $|V_{ts}|^2/|V_{td}|^2$, 
one has to use a value that is not contaminated by new physics.
Stated differently, one needs a measurement of the angle $\alpha^{\sm}$ or
$\gamma^{\sm}$ from pure SM processes. Various strategies for a clean
determination of these angles have been proposed, see
Ref.~\cite{gamma-th}, and are expected to yield stringent constraints in
the near future. For the time being, however, one has to resort to a
different method and exploit the very basic fact that a triangle is
completely determined by three parameters, which in our case are the
base, of length 1, 
the left side, which is determined by $|V_{ub}/V_{cb}|$, and the
angle $\beta^{\sm}$ between the base and the right side.
The essential assumptions that enter here are (i) that the determination
of $|V_{cb}|$ and $|V_{ub}|$ from semileptonic decays is free of new
physics, which is a model-independent assumption as these are
tree-processes, and that (ii) $\beta$ as measured from $B_d\to  J/\psi K_S$
is actually 
$\beta^{\sm}$ -- which, as mentioned above, is indeed the case in many SUSY
models, but is a more model-dependent statement than (i).
Using 
\bea
\sin 2\beta&=& 0.736 \pm 0.049\quad \cite{BaBar1,Belle1}\\
|V_{ub}/V_{cb}|&=& 0.090\pm0.025\quad \cite{PDG},
\eea
one obtains an allowed region for the position of the apex of the unitarity
triangle which is shown as shaded area in Fig.~\ref{fig:1}(a). The
allowed values of $\gamma^{\sm}$ are 
$45^{\circ}< \gamma^{\sm}< 100^{\circ}$.
$|V_{ts}/V_{td}|$ can be read off the figure as a function of $\gamma^{\sm}$
from the right side of the triangle and translated into an allowed
region for $\dM^{\sm}$ as shown in Fig.~\ref{fig:1}(b), where we also
include the error from $B_{B_s}f^2_{B_{s}}/(B_{B_d}f^2_{B_{d}})$. 
As can be seen from this figure, the current experimental bound 
$\Delta M_{s}>13\mbox{ps}^{-1}$ \cite{PDG} 
does not yet exclude any value of 
$\gamma^{\sm}$ between $45^{\circ}$ and $100^{\circ}$. 

Let us now turn to $\Delta \Gamma^{\sm}_{s}$. A recent estimate 
including NLO QCD corrections 
and lattice results for the hadronic parameters yields \cite{dG} 
\be
\frac{\dG^{\sm}}{\Gamma^{tot}_{s}}=(0.12\pm 0.06). \label{eq:23}
\ee
At present, there is no experimental bound.

\subsection{Observability of the $B_s^0-\overline{B_s}^0$ Oscillation}

A convenient measure of the frequency of the oscillation is the 
parameter $x_s$, defined as
$$
x_s\equiv \frac{\dM}{\Gamma_{B_s}}\,;
$$
$x_s$ indicates the observability of the oscillation, which is
governed by $\sin (x_s t/\tau_s)$; it is evident that the experimental
resolution of
rapid oscillations with $x_s\gg 1$ is extremely difficult. 
The current experimental lower bound is $x_s>19$;
recent studies of the experimental reach of the BTeV \cite{BTeVr} 
and the LHC \cite{LHCr}
experiments indicate that $x_s$ can be measured up to values $x_s
\approx 90$ 
(note that the corresponding parameter in the $B_d$
system, $x_d$ has been measured to be 0.73). 
The performance of ATLAS, CMS and LHCb in analysing 
$B_s\to J/\psi \phi$ has also been studied, which allows the
determination of the correlation between 
the new physics mixing phase $\sin 2\beta_s$ and the frequency $x_s$ 
\cite{LHCr}. 
Although the sensitivity to $\sin 2\beta_s$ gets worse as $x_s$ increases,  
values of $\sin 2\beta_s$ as 
small as $\mathcal{O}(10^{-2})$ are within experimental reach for
moderate $x_s<40$. 

Let us now discuss the correlation between 
$2\beta_s$ and $x_s$ in terms of contributions from beyond SM. 
\begin{figure}[t]\vspace{-0.5cm}\begin{center}
\psfrag{R=0.3}[l][l][0.8]{$|R|=0.3$}
\psfrag{R=0.5}[l][l][0.8]{$|R|=0.5$}
\psfrag{R=0.8}[l][l][0.8]{$|R|=0.8$}
\psfrag{R=1}[l][l][0.8]{$|R|=1$}
\psfrag{R=2}[l][l][0.8]{$|R|=2$}
\psfrag{R=3}[l][l][0.8]{$|R|=3$}
\psfrag{R=4}[l][l][0.8]{$|R|=4$}
\psfrag{R=5}[l][l][0.8]{$|R|=5$}
\psfrag{0to}[l][l][0.8]{$\arg [R]=0\to \pi$}
\psfrag{pito}[l][l][0.8]{$\arg [R]=\pi \to 2\pi$}
\psfrag{x}[l][l][1.1]{$x_s$}
\psfrag{tb}[l][l][1.1]{$2\beta_s$}
\psfrag{g}[l][l][1.1]{$\dG/\dG^{\sm}$}
\psfrag{argR}[l][l][1.1]{$\arg [R]$}
\includegraphics[width=16cm]{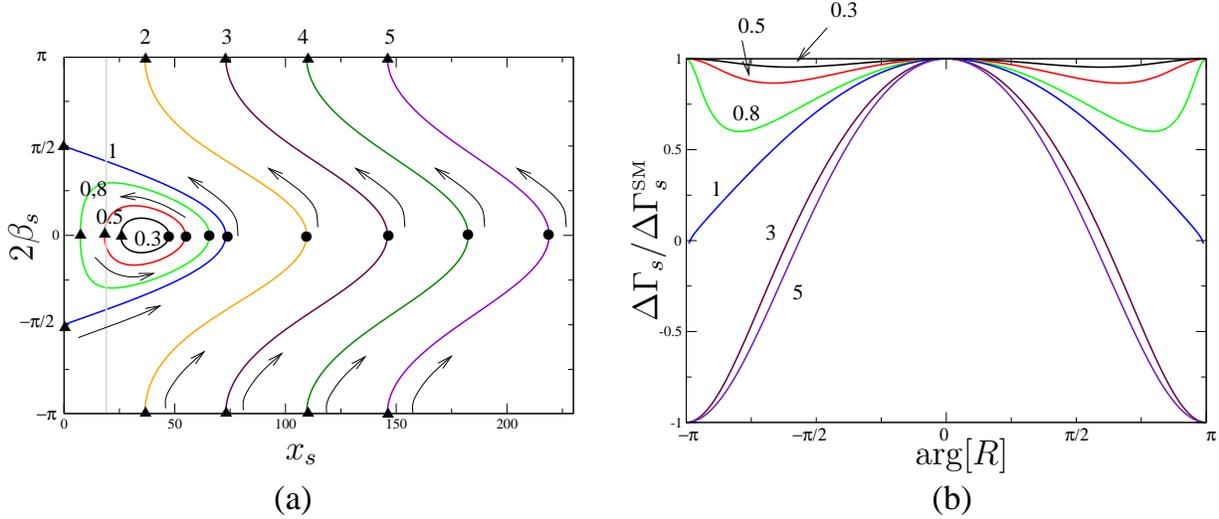}
\caption{(a) Correlation between $x_s$ and $2\beta_s$ for 
$|R|\in\{0.3, 0.5, 0.8, 1, 3, 5\}$ and $\arg R\in[0, 2\pi]$, where $R$ 
parametrises
the new physics contributions to $M_{12}$, Eqs.~(\ref{eq:25}), (\ref{eq:26}). 
The numbers in the figure represent the values of $|R|$ 
and the circles and triangles 
indicate $\arg R =0$ and $\pi$, respectively. The value of 
$\arg R$ increases in the direction of the arrow.  
The perpendicular line is the current experimental lower bound of $x_s$. 
(b) New physics in 
$\dG$. The numbers in the figure represent the value of $|R|$. 
$|\dG|$ is always reduced by new physics 
and can even become zero.}
\label{fig:2}
\vspace{-0.5cm}
\end{center}\end{figure}
For later convenience, we parametrise the new physics contributions as
\be R\equiv \frac{M_{12}^{\mbox{\tiny NP}}}{M_{12}^{\sm}}\,,\label{eq:25}\ee
which implies
\be 2\beta_s=\arg [1+R], \ \ \ \ \ x_s=
\frac{\Delta M_s^{\sm}}{\Gamma_{s}}\left|1+R\right|. 
\label{eq:26}\ee
In Fig.~\ref{fig:2}(a) we plot the correlation between $2\beta_s$
and $x_s$ for different values of $|R|\in$\{0.3, 0.5, 0.8, 1, 3, 5\} 
varying the
phase $\arg R$ between 0 and $2\pi$. 
The value of $\Delta M_s^{\sm}$ is chosen to be $25{\rm ps}^{-1}$. 
The figure shows that the current experimental bound on 
$x_s$ has already excluded some phase region for $0.5<|R|<1$.  
In view of the limitation of the experimental resolution, $x_s<90$, it
is clear that new physics can only be resolved if it is not too large,
i.e.\  $|R|<4$. 
As for the mixing phase, $2\beta_s$, small $|R|\ll 1$ will result in
small $2\beta_s$ that cannot be distinguished from the SM expectation,
unless $\arg R$ is very close to zero or $\pi$. 
For large SUSY  contributions $|R|>1$, on the other hand, 
$\sin 2\beta_s \simeq 1$ is very possible. 

Let us now discuss new physics effects on $\Delta \Gamma_{s}$. 
As discussed in \cite{DDLR,Gros}, $\Delta\Gamma_s$ is always reduced
by  new physics due to the factor $\cos 2\beta_s$ in Eq.~(\ref{xyz}). 
In Fig.~\ref{fig:2}(b), we plot $\dG/\dG^{\sm}$ in terms of $\arg R$ 
for different values of $|R|$. 
As can be seen from this figure, $\Delta\Gamma_s$ can even become 
zero for large values of $|R|$ and $\arg R=\pm \pi /2 $. 
\begin{figure}[ht]\vspace{-0cm}\begin{center}
\psfrag{z}[l][l]{zoom in}
\psfrag{t}[l][l]{$t\,$[ps]}
\psfrag{a}[l][l]{$S_{J/\psi \phi}$}
\includegraphics[width=15cm]{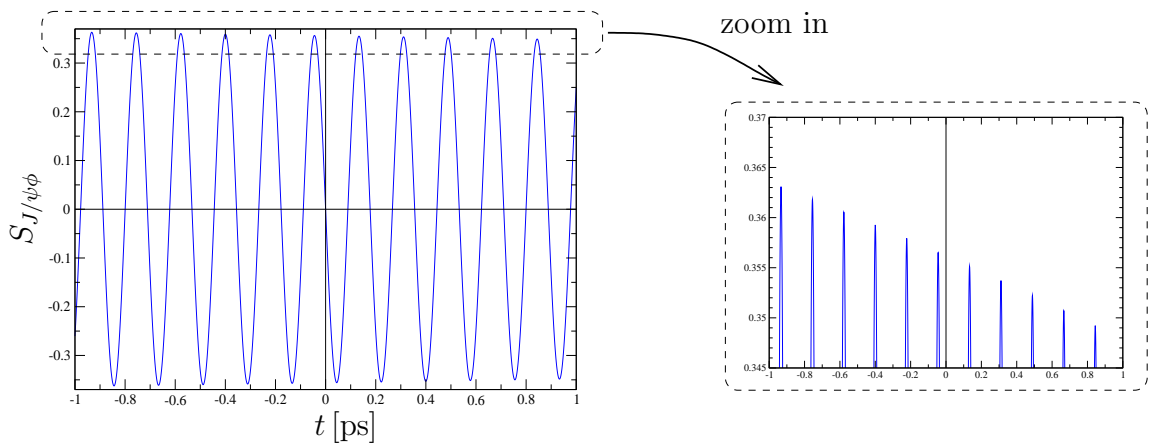}
\caption{The time-dependent asymmetry of $B_s\to J/\psi \phi$ acc.\ to
Eq.~(\ref{eq:1}); parameters as given in the text.} 
\label{fig:3}
\vspace{-0.5cm}
\end{center}\end{figure}

Finally, let us discuss the effect of $\Delta\Gamma_{s}$ on the 
time-dependent asymmetry Eq.~(\ref{eq:1}).  In Fig.~\ref{fig:3} 
we show the time-dependent asymmetry of  $B_s\to J/\psi \phi$ for   
the parameter set $\dM = 25\, {\rm ps}^{-1}$, 
$\dG^{\sm}/\Gamma_{s}^{tot} = 0.12 $, $D=0.33$, $|R|=1$ and $\arg R=\pi/2$.  
Note that the maxima of the $\sin\Delta M_s t$ curve slowly decreases 
with $t$, which is the effect of the denominator of Eq.~(\ref{eq:1}). 
Although this effect is rather small, it may be used to determine
$\Delta\Gamma_{s}$ once experimental data become available in
a sufficiently large range of $t$.


%
\section{\large \bf\boldmath 
SUSY Contributions to $B_s$ Mixing}
The mass difference in the $B_s$
system and the time-dependent asymmetry $S_{J/\psi \phi}$ depend
essentially on $M_{12}$ which can be computed from the effective
$\Delta B=2$ Hamiltonian $H_{\mathrm{eff}}^{\Delta B=2}$.
In supersymmetric theories
$H_{\mathrm{eff}}^{\Delta B=2}$ is generated by the SM box diagrams
with $W$ exchange and box diagrams mediated by 
charged Higgs, neutralino, gluino and chargino exchange. The
Higgs contributions are suppressed by the quark masses and can be neglected.
Neutralino diagrams are also heavily suppressed compared
to the gluino and chargino ones, due to the electroweak
neutral couplings to fermion and sfermions.
Thus, the
$B^0$--$\bar B^0$ transition matrix element is to good accuracy given by
\be
M_{12} = M_{12}^{\mathrm{SM}} +  M_{12}^{\tilde{g}}+ 
M_{12}^{\tilde{\chi}^+},
\ee
where $M_{12}^{\mathrm{SM}}$, $M_{12}^{\tilde{g}}$ and 
$M_{12}^{\tilde{\chi}^+}$
indicate the SM, gluino and chargino contributions, respectively. 
The SM contribution
is known at NLO accuracy in QCD \cite{buras1} and is given by
\begin{equation}
M_{12}^{\sm}= \left(\frac{G_F}{4\pi}\right)^2(V_{tb}^*V_{ts})^2
S_0(x_t) \eta_{2B}[\alpha_s(\mu)]^{-6/23}\left[1+\frac{\alpha_s(\mu)}
{4\pi}J_5\right] \left(-\frac{4}{3}m_{B_s}f_{B_s}^2B_1(\mu)\right),
\end{equation}
where $S_0(x_t)$ is given by
\begin{equation}
S_0(x_t)=\frac{4x_t-11x_t^2+x_t^3}{4(1-x_t)^2}-\frac{3x_t^3\ln x_t}{2(1-x_t)^3}
\end{equation}
with $x_t=(m_t/m_W)^2$. Contributions from virtual $u$ and $c$ quarks
are suppressed by the GIM mechanism.
The short-distance QCD corrections are encoded in $\eta_{2B}$ and 
$J_5$, with $\eta_{2B}=0.551$ and $J_5=1.627$ \cite{buras1}.

Including gluino and 
chargino exchanges, $H_{\mathrm{eff}}^{\Delta B=2}$ takes the form
\be
H_{\mathrm{eff}}^{\Delta B=2} = \sum_{i=1}^5 C_i(\mu) Q_{i}(\mu) +
\sum_{i=1}^3 \tilde{C}_i(\mu) \tilde{Q}_i(\mu) + h.c. ,  
\ee 
where $C_i(\mu)$, $\tilde{C}_i(\mu)$, $Q_i(\mu)$ and $\tilde{Q}_i(\mu)$ are 
the Wilson-coefficients and effective 
operators, respectively, normalised at the scale $\mu$, with
\bea
Q_1 &=& \bar{s}^{\alpha}_L \gamma_{\mu} b_L^{\alpha}~ \bar{s}^{\beta}_L 
\gamma^{\mu} b_L^{\beta},\nonumber\\
Q_2 &=& \bar{s}^{\alpha}_R b_L^{\alpha}~ \bar{s}^{\beta}_R 
b_L^{\beta},\nonumber\\
Q_3 &=& \bar{s}^{\alpha}_R b_L^{\beta}~ \bar{s}^{\beta}_R 
b_L^{\alpha},\nonumber\\
Q_4 &=& \bar{s}^{\alpha}_R b_L^{\alpha}~ \bar{s}^{\beta}_L 
b_R^{\beta},\nonumber\\
Q_5 &=& \bar{s}^{\alpha}_R b_L^{\beta}~ \bar{s}^{\beta}_L 
b_R^{\alpha}.
\eea
The operators $\tilde{Q}_{1,2,3}$ are obtained from $Q_{1,2,3}$ by
exchanging $L \leftrightarrow R$.

In MIA, the gluino contributions to the
 Wilson-coefficients  at the SUSY scale $M_S$ are given by
\cite{masiero}
\bea
C_1^{\tilde{g}}(M_S)\!&=&\!-\frac{\alpha_s^2}{216 m_{\tilde{q}}^2} 
\left[ 24 x f_6(x) + 66 \tilde{f}_6(x) \right] (\delta_{23}^d)^2_{LL} \\
C_2^{\tilde{g}}(M_S)\! &=&\!-\frac{\alpha_s^2}{216 m_{\tilde{q}}^2} 
204 x f_6(x) (\delta_{23}^d)^2_{RL} \\
C_3^{\tilde{g}}(M_S)\! &=&\!-\frac{\alpha_s^2}{216 m_{\tilde{q}}^2} 36  x f_6(x) 
(\delta_{23}^d)^2_{RL}  \\
C_4^{\tilde{g}}(M_S)\! &=&\!-\frac{\alpha_s^2}{216 m_{\tilde{q}}^2} \left\{
\left[ 504 x f_6(x) -72
 \tilde{f}_6(x) \right] (\delta_{23}^d)_{LL}(\delta_{23}^d)_{RR} -132 
\tilde{f}_6(x) (\delta^d_{23})_{LR} (\delta^d_{23})_{RL} \right\} \\
C_5^{\tilde{g}}(M_S)\! &=&\!-\frac{\alpha_s^2}{216 m_{\tilde{q}}^2} \left\{
\left[ 24 x f_6(x) +120 \tilde{f}_6(x) \right] (\delta_{23}^d)_{LL}
(\delta_{23}^d)_{RR} -180 \tilde{f}_6(x) (\delta^d_{23})_{LR} 
(\delta^d_{23})_{RL} \right\}.
\eea
where $x=m_{\tilde{g}}^2/m_{\tilde{q}}^2$ and $m_{\tilde{q}}$ is the average 
down squark mass. 
Explicit expressions for $f_6(x)$ and $\tilde{f}_6(x)$ 
can be found in \cite{masiero}. The Wilson-coefficients 
$\tilde{C}_{1,2,3}$ are obtained by 
interchanging $L\leftrightarrow R$ in the mass insertions appearing 
in $C_{1,2,3}$.
Note that the coefficient of the mass insertion
$(\delta^d_{23})_{LL} (\delta^d_{23})_{RR}$ in $C_4^{\tilde{g}}$ 
is much larger than the coefficients of the other mass insertions,
which renders $\Delta M_{B_s}$ and $S_{J/\psi \phi}$ very sensitive to
these insertions.

The chargino contributions to the relevant Wilson-coefficients, at leading 
order in MIA, next-to-leading order in the
Wolfenstein parameter $\lambda$ and including the 
effects of a potentially light right-stop, are given by \cite{GK}
\bea
C_1^{\tilde{\chi}^+}(M_S) &=& \frac{\alpha^2}{48 m_{\tilde{q}}^2}
\sum_{i,j} 
\Big\{ \vert V_{i1}\vert^2 \vert V_{j1} 
\vert^2 \Big[(\delta^u_{32})_{LL}^2 + 2 \lambda (\delta^u_{31})_{LL} 
(\delta^u_{32})_{LL} \Big]
L_{2}(x_i,x_j)\nonumber\\
&&-2 Y_t \vert V_{i1} \vert^2 V_{j1} V_{j2}^*
\Big[(\delta^u_{32})_{LL} 
(\delta^u_{32})_{RL}
+ \lambda (\delta^u_{32})_{LL} (\delta^u_{31})_{RL} \Big] R_2(x_i, x_j,z)
\nonumber\\
&&+ Y_t^2 V_{i1} V_{i2}^* V_{j1} V_{j2}^* \Big[(\delta^u_{32})_{RL}^2
  + 2 \lambda 
(\delta^u_{32})_{RL} (\delta^u_{31})_{RL} \Big]\tilde{R}_2(x_i,x_j,z)\Big\},\\
C_3^{\tilde{\chi}^+}(M_S) & = & \frac{\alpha^2}{12 m_{\tilde{q}}^2}
\sum_{i,j} U_{i2} U_{j2} V_{j1} V_{i1} 
\Big[ (\delta^u_{32})_{LL}^2 + 2 \lambda (\delta^u_{32})_{LL} 
(\delta^u_{31})_{LL} \Big] L_{0}(x_i,x_j),
\eea
where $x_i=m^2_{\chi_i^+}/m^2_{\tilde{q}}$, 
$z=m^2_{\tilde{t}_R}/m^2_{\tilde{q}}$ and the functions
$R_2(x,y,z)$, $\tilde{R}_2(x,y,z)$, $L_0(x,y)$ and $L_2(x,y)$ are given in 
\cite{GK}. $U_{i,j}$ and $V_{i,j}$ are the unitary matrices that 
diagonalise the chargino mass matrix and $Y_t$ is the top 
Yukawa coupling (for more details, see \cite{GK}).  
Note that, neglecting the effect of the Yukawa couplings 
of the light quarks, the chargino contributions to 
$C_4$ and $C_5$ are negligible and that charginos do not contribute to 
$C_2(M_S)$ and $\tilde{C}_2(M_S)$ due to the colour structure of the
diagrams;
nonzero values at lower scales are however induced by QCD mixing effects.

To obtain the Wilson-coefficients at the scale $\mu\sim m_b$ one has
to solve the corresponding renormalisation group equations, which 
to LO accuracy was done in Ref.~\cite{BdBd-th0}, with the result
\be
C_r(\mu) = \sum_{i} \sum_{s} (b_i^{(r,s)} + \eta c_i^{(r,s)}) 
\eta^{a_i} C_{s}(M_S),
\ee
where $\eta = \alpha_s(M_S)/\alpha_s(\mu)$. The coefficients $b_i^{(r,s)}$,
$c_i^{(r,s)}$ and $a_i$ are given in Ref.~\cite{BdBd-th0}. 

In order to calculate $M_{12}$, we also need the matrix elements of
the effective operators $Q_i$ and $\tilde{Q}_i$ over $B_s$ meson
states. As usual, the matrix elements are expressed in terms of the
decay constant $f_{B_s}$, using the vacuum insertion approximation;
terms neglected in this approximation are included in a bag factor 
$B_i$ which is expected to be of order one.
One has
\bea
\langle \overline{B_s}^0 \vert Q_1 \vert B_s^0 \rangle &\equiv& -\frac{1}{3} m_{B_s} f_{B_s}^2 B_1(\mu),\\
\langle \overline{B_s}^0\vert Q_2 \vert B_s^0 \rangle &\equiv& \frac{5}{24} \left(\frac{m_{B_s}}{m_b(\mu)+
m_s(\mu)}\right)^2 m_{B_s} f_{B_s}^2 B_2(\mu),\\
\langle \overline{B_s}^0 \vert Q_3 \vert B_s^0 \rangle &\equiv& - \frac{1}{24} \left(\frac{m_{B_s}}{m_b(\mu)+
m_s(\mu)}\right)^2 m_{B_s} f_{B_s}^2 B_3(\mu),\\
\langle \overline{B_s}^0 \vert Q_4 \vert B_s^0 \rangle &\equiv& - \frac{1}{4} \left(\frac{m_{B_s}}{m_b(\mu)+
m_s(\mu)}\right)^2 m_{B_s} f_{B_s}^2 B_4(\mu),\\
\langle \overline{B_s}^0 \vert Q_5 \vert B_s^0 \rangle &\equiv& - \frac{1}{12} \left(\frac{m_{B_s}}{m_b(\mu)+
m_s(\mu)}\right)^2 m_{B_s} f_{B_s}^2 B_5(\mu);
\eea
the matrix elements of $\tilde{Q}_i$ are the same as for $Q_i$.
The hadronic parameters $f_{B_s}$ and $B_i$ have been calculated on
the lattice, yielding \cite{lat2}\footnote{
The overall sign is different from the one in \cite{lat2}, which is 
due to the different sign choice of the CP transformation; we chose 
{\bf CP}$|P^0\rangle = +|\overline{P}^0 \rangle$. }
$B_1(m_b) = 0.86(2)(^{+5}_{-4})$, $B_2(m_b)=0.83(2)(4),\ B_3(m_b) = 
1.03(4)(9),\ 
B_4(m_b)=1.17(2)(^{+5}_{-7})$, and $B_5(m_b)=1.94(3)(^{+23}_{-7})$; as
we shall see in the next section, we do not need a numerical value for
$f_{B_s}$.

\section{{\large \bf Numerical Analysis and Discussion}}
Let us now proceed to the numerical analysis of the impact of SUSY
effects on $\Delta M_{B_s}$ and $\sin 2 \beta_s$, which is most
conveniently done by studying the ratio $R$, Eq.~(\ref{eq:25}), 
of intrinsically supersymmetric to SM contributions to $M_{12}$. 
We start with the gluino contributions, which, as discussed in 
the previous section, depend on the average down 
squark mass and on the ratio $x=(m_{\tilde{g}}/m_{\tilde{q}})^2$. 
In terms of the mass-insertion parameters $\delta^d_{23}$, 
$R$ can be written as
\bea
R_{\tilde g} \equiv 
\frac{M_{12}^{\tilde{g}}}{M_{12}^{\mathrm{SM}}} &\simeq& a_1(m_{\tilde{q}},x) 
\left[(\delta^d_{23})_{LL}^2 + (\delta^d_{23})_{RR}^2 \right] + 
a_2(m_{\tilde{q}},x) \left[(\delta^d_{23})_{LR}^2 + (\delta^d_{23})_{RL}^2 
\right] \nonumber\\
&& +a_3(m_{\tilde{q}},x) \left[(\delta^d_{23})_{LR} (\delta^d_{23})_{RL}
\right] + a_4(m_{\tilde{q}},x) \left[(\delta^d_{23})_{LL} 
(\delta^d_{23})_{RR}\right] \label{eq:Rg}
\eea
\begin{figure}[t]
\begin{center}
\psfrag{a}[l][l]{$a_i(m_{\tilde{q}},x)$}
\psfrag{a1}[l][l]{$a_1$}
\psfrag{a2}[l][l]{$a_2$}
\psfrag{a3}[l][l]{$a_3$}
\psfrag{a4}[l][l]{$a_4$}
\psfrag{x}[l][l]{$x=(m_{\tilde{g}}/m_{\tilde{q}})^2$}
\epsfig{file=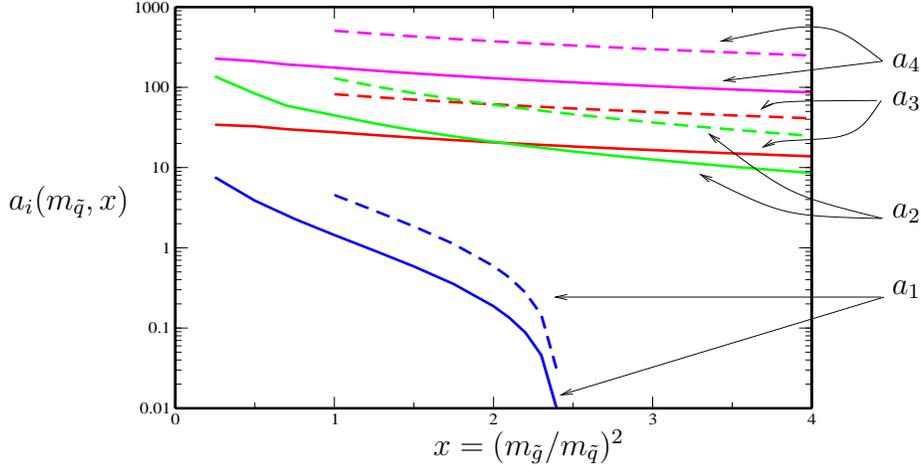,width=12cm,height=6cm}\\
\caption[]{$a_i(m_{\tilde{q}},x)$ defined in Eq.~(\ref{eq:Rg}) 
as function of 
$x=(m_{\tilde{g}}/m_{\tilde{q}})^2$ for $m_{\tilde{q}}=500\,$GeV (solid lines) 
and 300 GeV (dashed lines).}
\label{fig:4}
\end{center}
\end{figure}
with $x=m_{\tilde g}^2/m_{\tilde q}^2$. The 
coefficients $a_i(m_{\tilde{q}},x)$ depend implicitly on the
Wilson-coefficients and matrix elements defined in the previous 
section. Let us pause here for a moment and consider what range of
values for $\delta^d_{23}$ we actually do expect. Although our analysis
is model-independent, we may nevertheless get some guidance for what
to expect by looking at
various SUSY models. For instance, with $m_{\tilde{q}}\sim m_{g}
\sim 500\,$GeV,
the minimal supergravity  model gives 
$(\delta^d_{23})_{LL} \simeq 0.009+ 0.001~ i$ and
$(\delta^d_{23})_{RR, LR, RL} \simeq 0$, while the SUSY SO(10) model
predicts $(\delta^d_{23})_{RR}\simeq 0.5 + 0.5~ i $ and
$(\delta^d_{23})_{LL, LR, RL} \simeq 0$ \cite{BdBd-th1}. 
Models with nonuniversal A-terms lead to
$(\delta^d_{23})_{LR}\simeq 0.002+ 0.005~ i$ and
$(\delta^d_{23})_{LL, RR, RL} \simeq 0$ \cite{Khalil}.
We thus see that, although this is not expected to be true in general,
a single mass insertion is dominant in many models.
This implies that, for  
$(\delta^d_{23})_{LL, RR}$ ($(\delta^d_{23})_{LR, LR}$)
dominated models, only the term proportional to
$a_1(m_{\tilde{q},x})$ ($a_2(m_{\tilde{q},x})$) contributes to $R$.
We would also like to mention that
$(\delta^d_{23})_{AB}$ is  already constrained by
$B(b\to s \gamma)$, which
yields $|(\delta^d_{23})_{LL, RR}|<1$ and
$|(\delta^d_{23})_{LR, RL}|<\mathcal{O}(10^{-2})$ \cite{bsgamma}.

Numerical results for the  $x$ dependence of 
$a_i(m_{\tilde{q}},x)$ are given in  Fig.~\ref{fig:4}, 
for two representative values of the down squark mass,  
$ m_{\tilde{q}}=\{300, 500\}$ GeV. 
In order to obtain this result, we have set $M_S=m_{\tilde{q}}$ and
used the following input parameters:
\bea
&&V_{ts}=0.0412, \quad m_t=(174 \pm
5)\,\mbox{GeV},\quad\alpha_s(M_Z)=0.119,\nonumber \\
&&m_b(m_b)=4.2\  \mbox{GeV},\quad 
\mu=m_b,\quad m_s(2\ \mbox{GeV})=(100\pm 20)\,\mbox{MeV}.\nonumber
\eea
The impact of the theoretical uncertainties of $m_t$ and $m_s$ on
$a_i$ is very small, and also the variation with $\mu\sim m_b$ does
not exceed a few percent. 
The main source of uncertainty of $a_i(m_{\tilde{q}},x)$ comes from the 
$B_i$ parameters: although the factor $B_1$ cancels in $a_1$, the
other $a_i$ carry a $\sim 20$\% uncertainty from $B_i/B_1$. Note that
$R_{\tilde g}$ is independent of $f_{B_s}$. 

Let us continue with the discussion of the results depicted in 
Fig.~\ref{fig:4}. The solid and dashed lines refer to  
$m_{\tilde{q}}=500\,$GeV and 300 GeV, respectively. 
We see that all $a_i$ are monotonically decreasing functions in $x$
and  
are by about a factor 3 larger for $m_{\tilde{q}}=300$ GeV 
than for $m_{\tilde{q}}=500\,$GeV. 
Note also that $a_1(m_{\tilde{q}},x)$ becomes 
negative for large values of $x$. 
It is also evident that $a_4(m_{\tilde{q}},x)$ is 
largest, in agreement with the remark in the previous section, so that
the dominant contribution to 
$B_s$ mixing through gluino exchange is expected to be due to 
$LL$ and $RR$ mass insertions.
Although $a_{2,3}(m_{\tilde{q}},x) \sim\mathcal{O}(10)$ are also
large, the constraint from $B(b \to s \gamma)$ on the helicity-flip 
mass insertions $(\delta^d_{23})_{LR,RL}$ renders their contributions to 
$B_s$ mixing negligible. 

As an explicit example for the relative size of the $a_i$, we choose 
$m_{\tilde{q}}=500$ GeV and $x=1$, which yields
\bea
R_{\tilde{g}}
(m_{\tilde{q}}=500 \mbox{GeV}, x=1)&\simeq& 1.44 \left[
(\delta^d_{23})_{LL}^2 + (\delta^d_{23})_{RR}^2 \right] + 27.57 \left[
(\delta^d_{23})_{LR}^2 + (\delta^d_{23})_{RL}^2 \right] 
\nonumber\\
&&{} -44.76 \left[(\delta^d_{23})_{LR} (\delta^d_{23})_{RL}\right] - 
175.79\left[(\delta^d_{23})_{LL} (\delta^d_{23})_{RR}\right]. 
\eea
Using the constraints from $b\to s\gamma$,  
$|(\delta^d_{23})_{LR(RL)}|< 10^{-2}$ and $|(\delta^d_{23})_{LL(RR)})|< 1$, 
it is evident that helicity-flipping mass insertions contribute
$\mathcal{O}(10^{-3})$ to $R_{\tilde{g}}$, whereas 
single $LL$ or $RR$ mass insertions 
can yield  $\mathcal{O}(1)$ contributions. 

In Sec.~2, we have already discussed the dependence of $\dM$ and 
$\sin 2\beta_s$ on $R$, cf.\ Fig.\ref{fig:2}(a). 
The constraint from $b\to s\gamma$ implies that
$LR$ and $RL$ mass insertions alone cannot generate a value of
$2\beta_s$ larger than $\sim\mathcal{O}(10^{-3})$, 
which is too small to be observed at the Tevatron or the LHC. 
$LL$ and $RR$ mass insertions, on the other hand, can result in
sizeable -- and measurable -- values of the $B_s$ mixing phase:
for instance, $(\delta^d_{23})_{LL}=1\times e^{i\pi/4}$ yields 
$\dM/\dM^{\sm}=1.75$ and $\sin 2\beta_s=0.82$, while 
for $(\delta^d_{23})_{LL}\simeq (\delta^d_{23})_{RR}= 
0.1\times e^{i\pi/10}$ one finds 
$\dM/\dM^{\sm}=1.12$ and $\sin 2\beta_s=-0.93$. 
Note 
that for the same mass insertion, {\it i.e.}  
$(\delta^d_{23})_{LL}=1\times e^{\pi/4}$, 
the smaller squark mass,   
$m_{\tilde{q}}=300$ GeV, accompanied by $x=1$ gives 
 about 3 times larger $|R|$, 
{\it i.e.} $|R|>4$, which is beyond the experimental  reach at the
LHC, as discussed in Sec.~2.

Let us now turn to the chargino contributions.
The chargino mediated processes depend on five relevant SUSY low energy 
parameters: $m_{\tilde{q}}$, $m_{\tilde{t}_R}$,  $M_2$, $\mu$ and $\tan \beta$. 
With $m_{\tilde{t}_R}=150\,$GeV, $m_{\tilde{q}}=200\,$GeV, 
$M_2=\mu =300\,$GeV and $\tan \beta=5$, we find
\bea
\frac{M_{12}^{\tilde{\chi}^+}}{M_{12}^{\mathrm{SM}}} &\simeq& 10^{-4} (\delta^u_{31})_{LL}  (\delta^u_{32})_{LL}
+ 2\times 10^{-4}  (\delta^u_{32})_{LL}^2 + 9.8 \times 10^{-8}  (\delta^u_{32})_{LL} (\delta^u_{31})_{RL}
\nonumber\\
	&&+2\times 10^{-7} (\delta^u_{32})_{LL} (\delta^u_{32})_{RL}
+2.4 \times 10^{-7} (\delta^u_{31})_{RL} (\delta^u_{32})_{RL} + 
5.4 \times 10^{-7} (\delta^u_{32})_{RL} ,\label{eq:44}
\eea
which is obviously much smaller than the gluino contribution. 
Even though the chargino contributions are very sensitive to the value of
$\tan\beta$, an increase of  $\tan \beta$ to 50 only entails an enhancement
of the the first two terms in (\ref{eq:44}) from $10^{-4}$ to
$10^{-2}$ -- still not large enough to distinguish 
$\dM$ and $\sin 2 \beta_s$ from the SM prediction.

Let us  finally discuss the implication of the experimental data of 
the CP asymmetry 
in the  $B_d \to \phi K_s$ process,  
$S_{\phi K_s}$. As the underlying quark-level process is a $b\to s$
transition, it is clear that this process is governed by the same mass 
insertions, $(\delta^d_{23})_{AB}$. Since a possible 
hint of new physics may already have been seen in this mode, it is very 
interesting to analyse the implications of the experimental data on 
$S_{\phi K_s}$ for $B_s$ mixing. 
Let us first recall the main result of the supersymmetric contributions 
to $S_{\phi K_S}$ previously obtained in Ref.~\cite{KK-phk}: 
the mixing CP asymmetry is given by 
\begin{eqnarray}
S_{\phi K_S} = \Frac{\sin 2 \beta + 2 R_{\phi} \cos \delta \sin(\theta_{\phi} + 2 \beta) + 
R_{\phi}^2 \sin (2 \theta_{\phi} + 2 \beta)}{1+ 2 R_{\phi} \cos \delta \cos\theta_{\phi} +
R_{\phi}^2},
\end{eqnarray}
where $\delta$ is the difference of the strong phase between SM and SUSY, 
but assumed to be $\delta=0$ in the following 
(see \cite{KK-etak} for a more detailed discussion). 
$R_{\phi}$ is the absolute value of the ratio between SM and SUSY decay 
amplitude
and $\theta_{\phi}$ is its phase, that is 
\be
R_{\phi} e^{i\theta_{\phi}}\equiv\left(\frac{A^{SUSY}}{A^{SM}}\right)_{\phi K_S}. 
\ee
For $m_{\tilde{g}}\simeq m_{\tilde{q}}=500~ \mathrm{GeV}$, we obtain 
\be
R_{\phi} e^{i\theta_{\phi}}
\simeq 0.23 (\delta_{LL}^d)_{23} +
97.4  (\delta_{LR}^d)_{23} + 97.4  (\delta_{RL}^d)_{23} + 0.23 (\delta_{RR}^d)_{23}. 
\label{aa1} 
\ee
Considering the same constraint from $b\to s\gamma$, we arrive at the 
conclusion that the $LR$ or $RL$ mass insertion gives the largest contribution
to $S_{\phi K_s}$ while the $LL$ or $RR$ contribution is subdominant. 
In Ref.~\cite{KK-phk}, 
we found that it is very difficult to get a negative $S_{\phi K_{S}}$ from 
$LL$ or $RR$ mass insertion dominated models without decreasing 
$m_{\tilde{q}}$. 
\begin{table}[t]
\begin{center}
\begin{tabular}{|c|c|c|c|c|c|c|}\hline \hline
\multicolumn{7}{|c|}{$m_{\tilde{q}}=m_{\tilde{g}}=500$\ GeV} \\ \hline \hline
\multicolumn{4}{|c|}{Mass insertion sets} & \multicolumn{3}{|c|}{Results} 
\\ \hline
$\delta_{LL(RR)}$ & $\delta_{RR(LL)}$ & $\delta_{LR(RL)}$ & $\delta_{RL(LR)}$ & 
$\Delta M_s$ [ps$^{-1}]$ &$\sin 2\beta_s$ & $S_{\phi K_S}$ 
\\ \hline 
$1\times e^{-i\pi /2}$ & 0  & 0  & 0 & 10.7 & 0 & \phantom{-}0.50 \\ \hline 
$1\times e^{-i\pi /4}$ & 0  & 0  & 0 & 43.5 & -0.82 & \phantom{-}0.59 \\ \hline 
0 & 0  & $0.01\times e^{-i\pi /2}$  & 0 & 24.9 & 0 & -0.36 \\ \hline 
0 & 0  & $0.01\times e^{-i\pi /4}$  & 0 & 25.0 & -$2.8\times 10^{-3}$ &\phantom{-}0.19 \\ 
\hline 
$1\times e^{-i\pi /2}$ & $1\times e^{-i\pi /2}$ & 0  & 0 & $4.39 \times 10^3$ &0 & 
\phantom{-}0.25 \\ \hline 
$0.1\times e^{-i\pi /4}$ & $0.1\times e^{-i\pi /4}$  & 0  & 0 & 50 & 0.87 & \phantom{-}0.70 \\ 
\hline \hline
\multicolumn{7}{|c|}{$m_{\tilde{q}}=m_{\tilde{g}}=300$\ GeV} \\ \hline \hline
\multicolumn{4}{|c|}{Mass insertion sets} & \multicolumn{3}{|c|}{Results} 
\\ \hline
$\delta_{LL(RR)}$ & $\delta_{RR(LL)}$ & $\delta_{LR(RL)}$ & $\delta_{RL(LR)}$ & 
$\Delta M_s$ [ps$^{-1}$] &$\sin 2\beta_s$ & $S_{\phi K_S}$ 
\\ \hline 
$1\times e^{-i\pi /2}$ & 0  & 0  & 0 & 87.6 & 0 & \phantom{-}0.05 \\ \hline 
$1\times e^{-i\pi /4}$ & 0  & 0  & 0 & 115 & -0.98 & \phantom{-}0.37 \\ \hline 
0 & 0  & $0.01\times e^{-i\pi /2}$  & 0 & 24.8 & 0 & -0.76 \\ \hline 
0 & 0  & $0.01\times e^{-i\pi /4}$  & 0 & 25.0 & -$8.3 \times 10^{-3}$ & -0.15 \\ 
\hline 
$1\times e^{-i\pi /2}$ & $1\times e^{-i\pi /2}$ & 0  & 0 & $1.26 \times 10^4$ &0 & 
-0.52 \\ \hline 
$0.1\times e^{-i\pi /4}$ & $0.1\times e^{-i\pi /4}$  & 0  & 0 & 128 & 0.98 & \phantom{-}0.65 \\ 
\hline \hline
\end{tabular}
\caption{Numerical results for $\Delta M_{B_s}$, $\sin 2\beta_s$ and 
$S_{\phi K_s}$ for some representative values of $(\delta^d_{32})_{AB}$ ($A,B=
L,R$) for $m_{\tilde{q}}= m_{\tilde{g}}\in\{300, 500\}\,$GeV.}
\end{center}
\end{table}

The most interesting result we would like to emphasize 
here is that 
$B_s$ mixing and $S_{\phi K_{S}}$ 
are dominated by 
different mass insertions: $LL, RR$ and  $LR, RL$, respectively. 
In Table 1, we present our results for $\Delta M_{B_s}$, $\sin 2 \beta_s$ and 
$S_{\phi K_s}$ for various sets of the mass insertions with  
$m_{\tilde{q}}=m_{\tilde{g}}=\{300 \mbox{GeV}, 500 \mbox{GeV}\}$  
\footnote{In this table, 
the phases are chosen to be negative so that $S_{\phi K_S}$ 
becomes less than $S_{J/\psi K_S}$ (see the more detailed discussion 
in \cite{KK-phk}).}. 
As we have mentioned above, the $LL$ and $RR$ mass 
insertions may lower the value of $S_{\phi K_s}$ and make it comparable to  
experiment if the SUSY 
masses are light enough. 
In this case, however, $\Delta M_s$ becomes so large that  it cannot
be resolved experimentally.
On the other hand, although
 $LR$ or $RL$ dominated models can explain the experimental data 
of $S_{\phi K_s}$
and also predict $\dM \sim \dM^{\sm}$, which is good news for
the  experimental side, 
in this case $\sin 2 \beta_s$ is too small to be observed. 
Thus, once the oscilation is seen with a large amplitude 
at the Tevatron or the LHC, {\em all models with a
single dominant mass insertion 
 will be excluded.}
If the $B_s$ oscillations are resolved experimentally with $x_s<90$, 
the only surviving models
predicting a negative 
$S_{\phi K_s}$ and an observable $\sin 2 \beta_s$ and $\Delta M_s$,
are  SUSY models
with combined mass insertions effects. An 
example of this class of models could result in, for instance, the following 
mass insertions $(\delta_{32}^d)_{AB}$: 
\bea
&& \vert (\delta_{23}^d)_{LL} \vert \simeq 0.02 , \nonumber\\
&& \vert (\delta_{23}^d)_{RR} \vert \simeq 0.5 , \nonumber\\
&& \vert (\delta_{23}^d)_{LR} \vert \simeq \vert (\delta_{32})_{RL} 
\vert \simeq 0.005 , \nonumber\\
&& \mathrm{arg}[(\delta_{23}^d)_{LL}] \simeq \mathrm{arg}[ (\delta_{23}^d)_{RR}] 
\simeq -\frac{\pi}{4} , \nonumber\\
&& \mathrm{arg}[(\delta_{23}^d)_{LR}] \simeq \mathrm{arg}[ (\delta_{23}^d)_{RL}] 
\simeq -\frac{\pi}{2} , \nonumber
\eea
which lead to:
\bea
&& \Delta M_s \simeq 40\, {\rm ps}^{-1},\nonumber\\
&& \sin 2 \beta_s \simeq 0.86, \nonumber\\
&& S_{\phi K_S} \simeq -0.7. \nonumber
\eea
Such nonuniversal soft SUSY breaking terms 
($LR$  and $RL$ of order $10^{-3}$ and large $RR$)
are possible  in  models derived from  string theory, as discussed in,
 for instance, 
Ref.~\cite{Khalil}.

\section{Conclusions}
We have studied supersymmetric contributions to $B_s$
mixing and the mixing-induced CP asymmetry of $B_s\to J/\psi \phi$ in
the mass insertion approximation, including constraints from other
$b\to s$ processes, in particular $b\to s\gamma$ and $B_d\to \phi K_s$. 
The SM predictions for these quantities are 
 $S_{J/\psi \phi}\simeq 10^{-2}$ and $\dM = (10 - 30)\,{\rm ps}^{-1}$, 
depending on the value of $\gamma$. 
We have shown that in SUSY these predictions can change quite
drastically, which is mainly due to gluino exchange contributions, 
whereas the chargino contributions to these processes are negligible.
We find that values 
$S_{J/\psi \phi}\simeq \mathcal{O}(1)$ and $\dM = (10-10^4)\,{\rm
  ps}^{-1}$
are quite possible. 
We also 
find that unlike their effects on the CP asymmetry of $B_d \to \phi K_s$, 
the mass insertions $(\delta_{23}^d)_{LR(RL)}$ do not provide significant 
contributions to these processes, whereas 
$(\delta_{23}^d)_{LL(RR)}$ imply a 
large $\dM$ and $\sin 2 \beta_s$. 
We have argued that a clean measurement of 
the $B^0_s-\overline{B}^0_s$ oscillation
and a significant deviation of $S_{\phi K_s}$ from $S_{J/\psi K_s}$ 
would exclude SUSY models with a single dominant mass insertion, 
which predict either small oscillation and negative $S_{\phi K_s}$ or large 
oscillation and $S_{\phi K_s} \simeq S_{J/\psi K_s}$.

\vspace{1cm}
\noindent
{\large \bf Acknowledgements}\\
We would like to thank  U. Egede and T. Nakada for providing 
useful information on the experimental reach of LHCb. 
This work was partly supported by the Belgian
Federal Office for Scientific, Technical and Cultural Affairs through the
Interuniversity Attraction Pole P5/27.

\end{document}